# Spin-torque switching mechanisms of perpendicular magnetic tunnel junctions nanopillars

Jamileh Beik Mohammadi[1,2] and Andrew D. Kent[1]

[1]Center for Quantum Phenomena, Department of Physics, New York University, New York, NY 10003, USA

[2]Department of Physics, Loyola University New Orleans, New Orleans, LA 70118, USA

## Abstract

Understanding the magnetization dynamics induced by spin-transfer torques in perpendicularly magnetized magnetic tunnel junction nanopillars and its dependence on material parameters is critical to optimizing device performance. Here we present a micromagnetic study of spin-torque switching in a disk-shaped element as a function of the free layer's exchange constant and disk diameter. The switching is shown to generally occur by: 1) growth of the magnetization precession amplitude in the element center; 2) an instability in which the reversing region moves to the disk edge, forming a magnetic domain wall; and 3) the motion of the domain wall across the element. For large diameters and small exchange, step 1 leads to a droplet with a fully reversed core that experiences a drift instability (step 2). While in the opposite case (small diameters and large exchange), the central region of the disk is not fully reversed before step 2 occurs. The origin of the micromagnetic structure is shown to be the disk's non-uniform demagnetization field. Faster, more coherence and energy efficient switching occur with larger exchange and smaller disk diameters, showing routes to increase device performance.

# Introduction

Spin transfer torque switching of magnetization is being utilized for magnetic random-access memory (STT-MRAM) where it enables a low power embedded non-volatile memory [1, 2, 3]. Such electronically controlled magnetic memory only requires power for reading and writing, not to retain information, and can be denser that static RAM (SRAM) because only one transistor is needed per memory cell. The basic device is a magnetic tunnel junction with perpendicularly magnetized ferromagnetic electrodes, one free to switch, the other fixed. The memory states are layers magnetizations aligned parallel and antiparallel and current pulses are used to switch between the states. Understanding the switching mechanisms is essential to advancing the technology. As such there have been a number of experimental studies, mainly time-resolved electrical measurements, which measure the change in junction conductance during switching events [4, 5, 6]. These types studies provide information on the spatially averaged response as they measure conductance averaged over the junction area. Spatially resolved experiments are possible but have not been applied to magnetic tunnel junctions yet; there have been notable spatially and time-resolved x-ray microscopy studies of switching in spin-valve nanopillars, which are all metallic structures (magnetic layers separated by non-magnetic metallic layers) [7].

Micromagnetic modeling is a powerful method to simulate magnetization dynamics and switching mechanisms and enables studying the effect of geometry and material parameters, which are difficult to do in experiment. Such studies can guide material and device development and highlight the key factors determining the switching time, speed and energy. It can also illustrate the magnetization dynamics important to the switching. Several studies have already been conducted. Visscher and coworkers examined the first instabilities that can lead to switching [8, 9] and related research examined thermally activated reversal in CoFeB/MgO nanostructures [10]. A more recent study examined the effect of element diameter on the switching dynamics [11].

In this article we consider the role of both element size and free layer exchange constant on the switching mechanism. Of particular interest in our study is that the exchange constant of free layers used in STT-MRAM devices can be greatly reduced from bulk values and be varied considerably by free layer composition [12]. The effect of this reduction in exchange constant on the reversal dynamics has not been directly investigated nor have the reversal pathways been explored systematically as a function of the exchange constant and disk diameter. Our results illustrate the zero-temperature magnetization reversal mechanisms and domain wall dynamics important to spin-transfer switching.

## Device geometry and material parameters

We studied disk shaped free layers characteristic of those being actively developed for STT-MRAM devices. We assume that the spin-torque is associated with a perpendicularly magnetized polarizing layer (labelled **P** in Fig. 1(c)) while the Oersted field from the current and the stray field from the polarizing layer are not considered.

The free layer used in STT-MRAM typically consists of CoFeB with a W insertion layer and interfaces on both sides to MgO, denoted a dual MgO composite free layer [13, 14]. The added interfaces (CoFeB/MgO and CoFeB/W) increase the perpendicular anisotropy. The magnetic parameters of this composite free layer are taken to be a saturation magnetization $M_s$=1209 kA/m, Gilbert damping $\alpha$ =0.015, and perpendicular magnetic anisotropy $K_u = 1118$ kJ/m³. The free layer thickness is taken to be the effective thickness of the free layer with a 0.3-nm W insertion layer, $t$=1.5 nm. The junction diameters $d = 10, 15, 30, 50,$ and 80 nm are studied with exchange constants of $A$ =4 and 8.5 pJ/m. These exchange values are those of a single CoFeB layer and a composite free layer with a 0.3 nm W insertion layer, respectively. All the above mentioned parameters such as $M_s$, $\alpha$, $K_u$, $A$, and $t$ are the experimental values for the pMTJ free layers studied in [12]. For comparison, we have also studied free layers with 19 pJ/m (as this value is reported for thick CoFeB films [15]). Given these parameters, the domain wall width, $\lambda = 2\sqrt{A/K_{\text{eff}}}$, and the critical diameter $d_c$, $d_c = \frac{16}{\pi}\sqrt{A/K_{\text{eff}}}$, the diameter below which the thermally activated switching is expected to be coherent, are listed in Table 1. Here, $K_{\text{eff}}$ is the effective perpendicular anisotropy that is calculated based on the sum of perpendicular uniaxial anisotropy field and demagnetizing field. Since demagnetizing field is size-dependent, $K_{\text{eff}}$ and therefore $\lambda$ and $d_c$ depends on the size of the junction [16].

Table 1: The domain wall width and the critical length for coherent switching, $d_c$, for three exchange constant values that are used in this paper. The $\lambda$ and the $d_c$ values reported in this table are for a 10-nm element where $K_{\text{eff}} = 419.5$ kJ/m³.

| $A$ [pJ/m] | $\lambda$ [nm] | $d_c$ [nm] |
|---|---|---|
| 4.0 | 6.1 | 14.5 |
| 8.5 | 8.1 | 21.2 |
| 19.0 | 13.3 | 31.7 |

## Methods

Switching of the free layer of a circular disk is modeled using mumax with a spin transfer torque [17]. The magnetization dynamics of the free layer is described by

$$\dot{\mathbf{m}} = -\gamma\mu_0 \mathbf{m} \times \mathbf{H}_{\text{eff}} + \alpha \mathbf{m} \times \dot{\mathbf{m}} + \boldsymbol{\tau}_{\text{STT}}. \quad (1)$$

Here $\dot{\mathbf{m}}$ is the time derivative of the magnetization vector **m**, $\gamma$ is the gyromagnetic ratio, $\mu_0$ is the vacuum permittivity, $\mathbf{H}_{\text{eff}}$ is the effective field, $\alpha$ is Gilbert damping parameter and $\boldsymbol{\tau}_{\text{STT}}$ is the spin-transfer torque

$$\tau_{STT} = \frac{\hbar}{2e}\frac{J}{M_s t} P\, \boldsymbol{m} \times (\boldsymbol{m} \times \boldsymbol{p}). \quad (2)$$

Here $\hbar$ is Planck's reduced constant, $e$ is the charge of electron, $J$ is the current density, P is the spin polarization, (for all simulations in this paper P =0.5) and $\boldsymbol{p}$ is a unit vector in direction of spin polarization **P** that we have set along the $z$ direction (perpendicular to the plane of the disk) as shown in Fig. 1(c). We note that this is zero temperature model of the switching dynamics and we kept the micromagnetic cell size small compared to the exchange length and the size of the junction.

For a perpendicular magnetic tunnel junction, the zero-temperature critical current density (obtained from a macrospin model [18]) is $J_{c0} = \frac{2e\alpha\mu_0 M_s H_k t}{\hbar P}$, where $H_k$ is the effective perpendicular anisotropy field (uniaxial anisotropy field minus the demagnetizing field) that depends on the size of the junction [16]. We varied the overdrive, that is the ratio $J/J_{c0}$, between 1.2 and 10 to study its effect on the switching dynamics and the switching speed. The reversal dynamics is studied through the time evolution of the average perpendicular magnetization, $m_z$. We define the switching time as the time when $m_z = 0$. In addition, snapshots of free layer magnetization are used to examine the micromagnetic instabilities associated with important pre- and post-switching time events.

## Results

In order to illustrate the effect of junction size on the spin torque switching dynamics we present the time evolution of $m_z$ for $d = 10$, 15, 30, 50, and 80 nm diameter elements with $A$ =8.5 pJ/m (Fig. 1(a)). The switching time increases as the diameter of the junction increases. Moreover, $m_z$ vs. time curves for larger junctions are not as smooth as that of a 10-nm element, which indicates that the reversal is less coherent for larger elements. Both effects, delayed and spatially incoherent switching for larger elements, are even more dramatic when the exchange constant of the free layer is reduced to 4 pJ/m. For large elements, a significant difference in the switching path is observed for these two cases ($A$ =4 and 8.5 pJ/m). Moreover, the switching time depends strongly on the strength of the exchange interactions even for junctions that are as small as 30 nm in diameter.

The exchange constant sets the length scale of spatial inhomogeneities of magnetization and determines the type of instabilities that occur during the spin torque switching process. Investigating the images of the z component of magnetization of the individual cells at different times allow us to examine the coherent/incoherent switching dynamics and observe the key features in the spin-torque dynamics. Table 2 shows key moments in the switching dynamics for intermediate and larger size junctions.

The images of the free layer confirm that the switching of small elements, e.g. 10-nm diameter free layer with A=8.5 pJ/m is coherent (see Fig. 1(a), the single color circles illustrates a spatially uniform magnetization at different stages in the switching). For intermediate-sized elements (for example, a 30-nm diameter element with A=8.5 pJ/m) the first fully reversed region ($m_z$=-1) is at the disk edge. While for A=4 pJ/m, the first fully reversed region is at the center as it is shown in magnetization images in Table 2.

As a polarized current passes through the free layer, the precession amplitude grows at the center of the disk due to larger magnetostatic field at the center that leads to a lower effective perpendicular anisotropy field in the center—that makes it easier to excite this region with a spin current [10]. Following this event, the first region (at the center or edge of the element depending on the size of the disk) switches at $t_1$. The second micromagnetic instability occurs when the region with the largest precession amplitude hits the edge at $t = t_2$ and domain wall(s) form across the element. Small elements (d smaller than a critical diameter $d_c$ [16]) experience semi-coherent spin torque switching, meaning that a 180° wall does not form. However, for intermediate element diameters, a 180° domain wall is formed across the element (at $t = t_{DW}$) that traverses the disk until the magnetization of the free layer reverses completely. We have observed that the movement of the domain wall is not at a constant speed, as expected in simple models. In large junctions, $d > 2d_c$, the increasing precession amplitude continues to the point that the magnetization in the center of the disk reverses completely and a precessing reversed magnetic droplet is formed, similar to the droplets predicted [19] and directly observed in spin-transfer nanocontacts [20]. The droplet tends to drift from the center of the element toward the element boundary, in analogy to the droplet drift instability observed in nanocontacts [21, 22]. We have found that, depending on the size of the element relative to the critical diameter, the switching time, $t_{SW}$, can be earlier or later then the onset of the second instability the leads to the formation of a domain wall that traverses the element. The sequence of the aforementioned events for $d > 2d_c$ and $d_c < d < 2d_c$ is illustrated in Table 2. At $t_1$ the first cell in the simulation switches (i.e. has $m_z$=-1). At $t_2$ the reversed droplet hits the element boundary. $t_{SW}$ is the time when the average z-component of magnetization is zero and $t_{DW}$ is when there is a domain wall in the element.

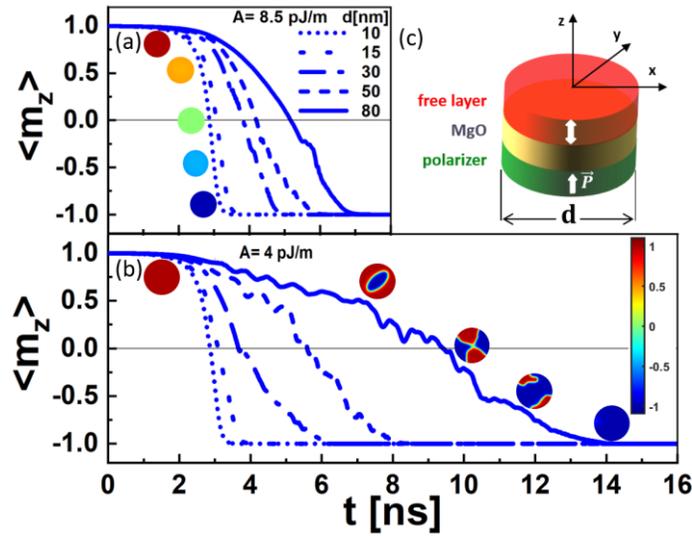

Figure 1: $m_z$ of the pMTJ free layer vs. time for 10, 15, 30, 50 and 80 nm diameter free layers with overdrive of 1.6 and exchange constant of of (a) $A$ =8.5 pJ/m and (b) $A = 4$ pJ/m. (c) Geometry of the pMTJ nanopillar studied. In (a) and (b), dark red, orange, green, light blue, and dark blue circles represent the $m_z$ profile of the 10-nm free layer with A=8.5 pJ/m at different times. A uniform color indicates that the magnetization switching is spatially coherent.

Inhomogeneous demagnetizing fields play an important role in the switching instabilities. To investigate the role of demagnetization fields, we have simulated spin torque switching of a 30-nm element with $A$ =4 and 8.5 pJ/m under the following two conditions. 1) When the demagnetizing field is present and the uniaxial anisotropy of the free layer is $K_u = 1118$ kJ/m³, and 2) when the demagnetizing field is not present and the uniaxial anisotropy is calculated using a thin disk approximation in which demagnetization tensor is diagonal with trace $N_{xx} + N_{yy} + N_{zz}$ = 1. In this case, we have set the uniaxial anisotropy to $K'_u = K_u - \frac{\mu_0 M_s^2}{4}(3N_{zz} - 1)$ [16]. For the second case, the reason to set the uniaxial anisotropy to $K_u'$ is to have the same effective perpendicular anisotropy as in condition 1. Figure 2 shows the time trace of $m_z$ for a 30-nm element with $A$ =4 pJ/m and $A = 8.5$ pJ/m under condition 1 and 2. An important observation is that in absence of the demagnetizing field, the switching is perfectly coherent, as the spatial profile of magnetization shows (Figure 2). Moreover, the curves are identical for the two exchange constant values, which indicates that the inhomogeneous demagnetizing field plays a critical role in determining the switching path and types of magnetic instabilities.

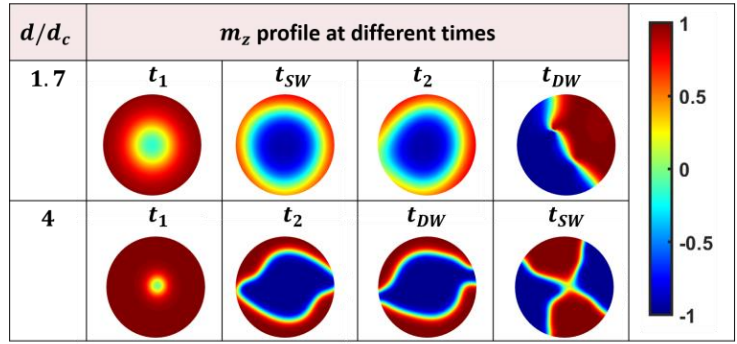

Table 2: pMTJ free layer $m_z$ profile at selected times for an intermediate size disk ( $d_c < d < 2d_c$, $d$ =30 nm, $d/d_c$ =1.7 (top row)) and a large disk ($d > 2d_c$, $d$ =80 nm, $d/d_c$ =4 (bottom row)). Exchange constant is 4 pJ/m. $t_1$ is the moment when a cell in the free layer element reverses. The magnetic droplet hits the edge at $t_2$ and $t_{DW}$ is when there is one (or two) 180-degree Bloch wall(s) that traverse the element. Time increases from left to right.

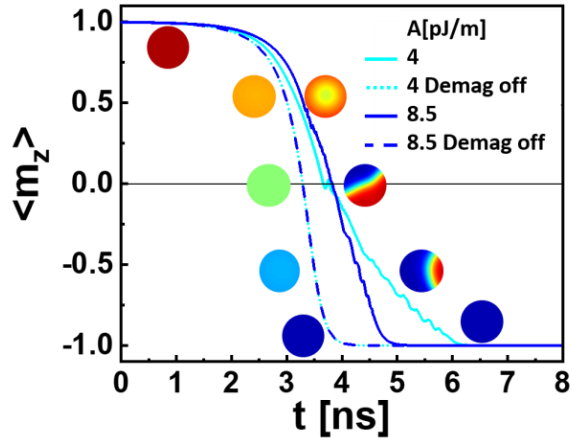

Figure 2: The solid (dashed) lines represent $m_z$ vs. time for a 30 nm pMTJ free layer with $A$ =4 and 8.5 pJ/m with (without) demagnetization field. The circles on the right (left) side of the curves show the image of $m_z$ for a 30-nm junction with A=8.5 in presence (absence) of the demagnetizing field. Solid color circles indicate that all cells have the same $m_z$ and color gradient of the right-side circles show the micromagnetic features associated with an incoherent reversal.

In order to investigate the effect of the overdrive on the switching dynamics, we have simulated a 30 nm diameter element with various exchange constant values, $A$ =2, 4, 8.5, and 19 pJ/m, and overdrive values from 1.2 to 10. Figure 4 shows $m_z$ vs. reduced time $\tau$ ($\tau$ is defined so that for each overdrive, $\tau$ =1 corresponds to the switching time for that overdrive) for overdrives between 2 and 10 for a 30-nm disk with $A$ =4 pJ/m (Fig. 4(a)), 8.5 pJ/m (Fig. 4(b)), and 19 pJ/m (Fig. 4(c)). For all exchange values, the effect of the overdrive on the switching path is negligible. As it can be seen in Fig. 4(d), when the exchange interaction is strong, that is when $A$ =8.5 and 19 pJ/m, the switching speed is well described by the macrospin model. However, for cases with weak exchange interaction, deviations from macrospin model are noticeable. As a result, the when exchange interaction is weak, i.e. $A$ =2 and 4 pJ/m, the slope of the switching speed vs. overdrive has a strong dependence on the exchange constant.

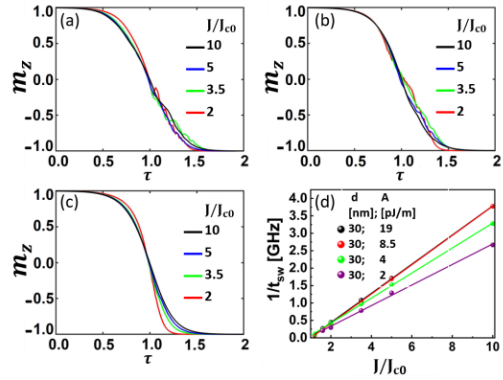

Figure 3: Free layer $m_z$ vs. reduced time for a 30 nm junction with (a) $A$ =4 pJ/m, (b) $A$ =8.5 pJ/m, and (c) $A$ =19 pJ/m for overdrive of 2, 3.5, 5, and 10. Switching speed vs. overdrive is plotted for the same junctions with various exchange constants in (d).

## Summary

In summary, we have simulated the spin torque switching of disk shaped pMTJ free layer to investigate the effect of the size of the element and the exchange interaction on the spin torque switching dynamics. For all junctions with different sizes and exchange constant values, the precession amplitude increases at the center of the element. A micromagnetic instability occurs when the region with growing precession amplitude hits the edge. At that moment, the junction is divided into two regions with a magnetization precession phase difference [9]. After this instability point, a domain wall (or a few domain walls) form across the element that leads to its full reversal. In addition, our results show the importance of the non-uniform demagnetizing fields in the magnetization nonuniformities, meaning that in absence of the demagnetizing field, spin-torque switching is observed to be coherent irrespective of the exchange constant or the size of the junction.

In addition to the switching dynamics, the switching time depends on the exchange constant of the free layer. For example, full reversal of a 30-nm element with A=4 pJ/m takes 6.90 ns which is 15 % longer than that of the same size element with A=8.5 pJ/m (when $J/J_{c0}$ =1.6). For an 80-nm element, the difference between the full reversal time for the same exchange values is more than 58 %. We find that the switching speed increases linearly with increasing overdrive irrespective of the exchange constant. However, for small exchange constant values (A≤4 pJ/m), the switching speed and the slope of the switching speed vs. overdrive depends on the exchange constant.

Our results show the effect of the exchange interaction on the spin torque switching dynamics and switching speed of pMTJ free layers and when the dynamics deviate from macrospin model predictions. This highlights the importance of investigating the magnetic properties of the free layer to better understand and predict the switching dynamics. It is particularly important to include the effect of stray fields of polarizer layer that can have stabilizing or destabilizing effects on the free layer magnetization depending on the initial magnetic configuration [23]. We conclude that a strong exchange interaction is required to ensure fast, energy efficient, and coherent spin torque switching of the free layer. Otherwise, delayed switching is expected as inhomogeneous magnetization dynamics occur. Experimentally, time-resolved measurements like [4, 5] and time-resolved microscopy experiments could provide more insights into the effect of the exchange interaction on the switching dynamics of pMTJ devices.

## Acknowledgements

This research was supported by Spin Memory, Inc. A.D.K. acknowledges support from the National Science Foundation under Grant No. DMR-1610416.